\documentclass[a4paper,11pt]{article}
\usepackage{jinstpub} 
\usepackage{lineno}
\usepackage{subfigure}
\usepackage{comment}
\usepackage[leftcaption]{sidecap}

\title{\boldmath Latest results from CUORE and prospects for CUPID}

\author{K. Zhao on behalf of the CUORE and CUPID collaborations}
\affiliation{Gran Sasso Science Institute, L’Aquila, Italy}
\affiliation{INFN Laboratori Nazionali del Gran Sasso, Assergi (AQ), Italy}

\emailAdd{kangkang.zhao@gssi.it}

\abstract{The search for neutrinoless double beta decay (0$\nu\beta\beta$) is fundamental for investigating lepton-number violation, probing new physics beyond the Standard Model, and determining whether neutrinos are Majorana particles. CUORE (Cryogenic Underground Observatory of Rare Events), a cryogenic bolometric experiment at LNGS, studies 0$\nu\beta\beta$ in $^{130}$Te using 988 TeO$_2$ crystals. It is a milestone of cryogenic detector arrays with a tonne-scale detector operated for more than 7~years below 15~mK. Since 2017, CUORE has accumulated over 2.9~tonne-years of exposure, achieving one of the leading 0$\nu\beta\beta$ limits and one of the most precise two-neutrino double beta decay (2$\nu\beta\beta$) half-life measurements thanks to a detailed background reconstruction across a broad energy range. Building on CUORE’s success, CUPID (CUORE Upgrade with Particle ID) aims to significantly enhance its 0$\nu\beta\beta$ discovery sensitivity to $10^{27}$ yr in $^{100}$Mo, covering the Inverted Hierarchy of neutrino masses. It will employ lithium molybdate (Li$_2$MoO$_4$) crystals enriched in $^{100}$Mo, alongside germanium light detectors with Neganov-Trofimov-Luke amplification, enabling simultaneous heat and light readout for enhanced background rejection. 
CUPID will reuse CUORE’s cryostat and infrastructure. Current efforts focus on detector performance validation, sensitivity studies, and finalizing the experimental design to maximize physics reach. This work presents the latest CUORE results and outlines the key milestones toward CUPID’s realization.}

\keywords{Neutrino-less double beta decay, Cryogenic detectors}

\begin{document}
\maketitle
\flushbottom
\section{Introduction}

Neutrinoless double-beta decay is a hypothetical process that violates lepton number by 2 units hence it would indicate the presence of new physics beyond the Standard Model. 
The search for 0$\nu\beta\beta$ probes the neutrino mass hierarchy and absolute mass of neutrinos, which becomes more and more attractive as the neutrino has been demonstrated with non-zero mass by the discovery of neutrino oscillation. On the other hand, the observation of 0$\nu\beta\beta$ would also possibly explain the origin of the matter-antimatter asymmetry in the universe via the mass generation mechanism and the leptogenesis mechanism.~\cite{dolinski2019neutrinoless, zel1981study} 

0$\nu\beta\beta$ can be experimentally searched for in isotopes where single beta decay is energetically forbidden, but double beta decay (2$\nu\beta\beta$) is possible. The distinguishable experimental signature of double beta decay is the sum of the two electron energies. 0$\nu\beta\beta$ would produce a mono-energetic peak at the Q-value, while 2$\nu\beta\beta$ exhibits a continuous spectrum with endpoint at the Q-value. 2$\nu\beta\beta$ has been observed for 11 nuclides with extremely low decay rates, corresponding to half-lives of 10$^{18}$-10$^{24}$ years~\cite{barabash2020precise}. To date, no confirmed observation of 0$\nu\beta\beta$ has been observed, giving the most stringent constraint of $T_{1/2}^{0\nu}\geq 10^{26}$ years on the lifetime~\cite{abe2023search}. The detection sensitivity of 0$\nu\beta\beta$ can be formulated as Eq.~\ref{eq:1},
\begin{equation}
    \mathcal{S}^{0\nu} \propto \epsilon\cdot\frac{\eta}{A}\cdot\sqrt{\frac{M\cdot t}{B\cdot \sigma}} 
    \label{eq:1}
\end{equation}
where $\epsilon$ is the total detection efficiency of 0$\nu\beta\beta$ events, $\eta$ and $A$ are the isotopic abundance and mole mass of the target nuclide respectively, $M\cdot t$ is the product of the target nuclide mass and livetime, i.e. the exposure of an experiment, $B$ is the background index (BI) in unit of ``counts/keV/kg/year'' and $\sigma$ is the energy resolution at the Q-value. Therefore, in order to implement highly sensitive experiment searching for 0$\nu\beta\beta$ decay, excellent energy resolution, ultra low background rate in the region of interest (ROI) and the scalability for large exposure are required. 

A sensitive experiment for 0$\nu\beta\beta$ decay searching lies mostly on the selection of the target nuclide and the detection technique. The favorable target nuclide requires a high Q-value for suppressing  natural radioactive background and a high natural abundance or low expenses of enrichment for large exposure. As well as the detection technique is supposed to be equipped with high detection efficiency, excellent energy resolution, scalability and particle discrimination capabilities for background rejection. However, it is only possible to build a competitive experiment that excels in few compatible aspects~\cite{giuliani2012neutrinoless}. For example, the LEGEND experiment utilizes high-purity Germanium (HPGe) detectors searching for the decay of $^{76}$Ge leveraging the excellent energy resolution and high material purity helpful on suppressing the contamination background~\cite{25tk-nctn}, but accompanied with relatively low Q-value and high material expenses. The KamLAND-Zen experiment employed scintillation detectors with large scalability has demonstrated the current leading result of 0$\nu\beta\beta$ decay searching even though it has a sub-optimal energy resolution~\cite{PhysRevLett.130.051801}. The NEXT experiment uses gaseous Xenon time projection chamber with tracking/topology capability to perform a great signal-background discrimination~\cite{Martinez-Vara:2023rft}. In the coming years, a number of large experiments will be deployed for 0$\nu\beta\beta$ decay searching~\cite{gomez2023search}, aiming to fully cover the inverted hierarchy region and explore part of the normal hierarchy region. 

\section{CUORE experiment and its latest results}

The Cryogenic Underground Observatory for Rare Events (CUORE) is an experiment utilizing the cryogenic bolometric technique to investigate the 0$\nu\beta\beta$ decay. CUORE is the first tonne-scale experiment based on bolometric technique, which stands on the shoulders of several successful previous pioneering experiments MiDBD, CUORICINO and CUORE-0, as well the detector development initiated by E. Fiorini~\cite{alessandrello1998preliminary, andreotti2011130te, alfonso2015search, brofferio2018contributed}. Cryogenic bolometers are phonon-mediated detectors, the energy deposition in an absorber by particle interaction is converted into phonons excitation: leading to the temperature rise and then read by the thermal sensor. Phonon is a kind of quasi-particle represents the quantized vibrational energy of crystal lattice. The average excitation energy of phonons at very low temperature can be reduced to the order of meV, according to the Debye model. Moreover, the largest fraction of deposited energy is finally in form of phonons/heat, so the bolometers could exhibit an excellent intrinsic energy resolution. Bolometers have to be operated at ultra-low temperature ($\sim$10~mK) to minimize the environmental thermal fluctuation and also make the heat capacity of the absorber low enough for an enhanced signal-to-noise (SNR) ratio. Cryogenic bolometers features a wide choice of absorber materials. The main drawback of bolometric technique is its slow time response. In fact, the decay time of cryogenic bolometers can range from~ms to~s depends on the absorber size and the type of thermal sensor, which makes bolometers are only suitable for low signal rate detection.
\begin{figure*}[!htb]
\begin{center}
\includegraphics[height=4.5cm]{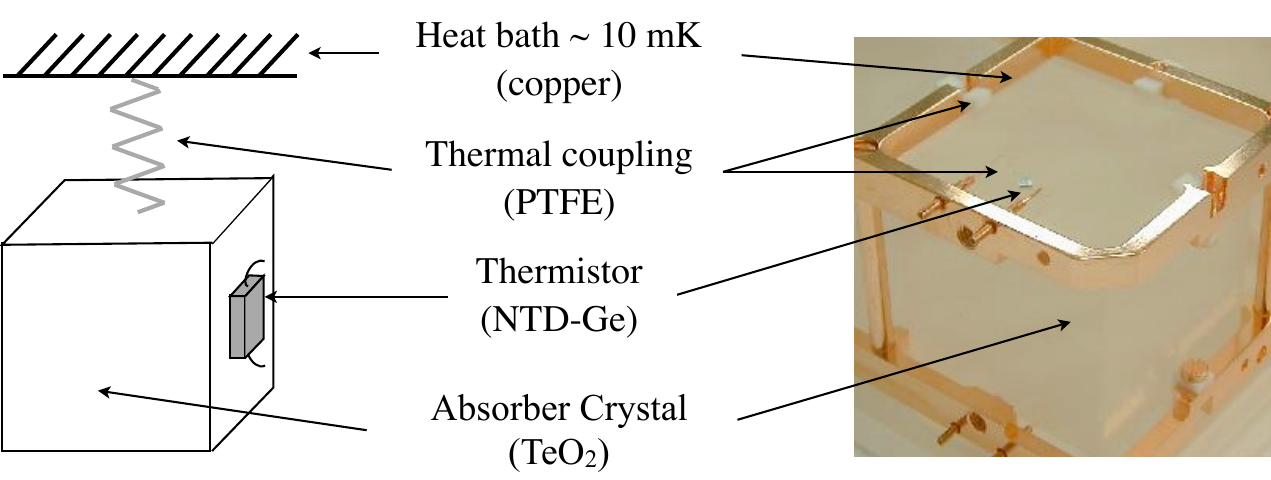}
\caption{\label{fig:bolo}Schematic of CUORE's bolometric detector module.}
\end{center}
\end{figure*}

The CUORE experiment is located at the Gran Sasso National Laboratories (LNGS) of INFN in central Italy. The CUORE detector consists of 988 TeO$_{2}$ crystals with a total mass of 742~kg (206~kg $^{130}$Te) to search for the 0$\nu\beta\beta$ decay of $^{130}$Te at the Q$_{\beta\beta}$ of 2527.5~keV. $^{130}$Te features the highest natural abundance among the double beta decay nuclides, so no enrichment process has been implemented in CUORE. The CUORE detector is hosted in a custom-designed cryogen-free cryostat and cooled down to its operational temperature of 15~mK~\cite{alduino2019cuore}. LNGS has ~3600 meters water equivalent of rock overburden: as a result the cosmic ray flux is reduced by 6 orders of magnitude compared to that of the surface. In addition, the CUORE experiment is equipped with an shielding against natural gamma radioactivity and neutrons. The average BI observed for CUORE is $\sim$1.42~counts/(keV$\cdot$kg$\cdot$yr) in the ROI~\cite{cuoreScience2025}. 

CUORE uses cubic TeO$_{2}$ crystal with side length 5~cm, the typical temperature rise is about 0.1~mK/MeV. Each crystal is thermally coupled a neutron transmutation doped (NTD) germanium thermistor by glue spots and used for the signal readout, as shown in Fig.~\ref{fig:bolo}. Each crystal is also instrumented with a silicon Joule heater to inject artificial heat pulses to stabilize the detector thermal gain. 
For CUORE’s detectors, we observed an average FWHM energy resolution of $\sim$7.54~keV at the 2615~keV calibration peak, which is second only to that of HPGe detectors~\cite{cuoreScience2025}.

\begin{figure*}[!htb]
\begin{center}
\includegraphics[height=5.0cm]{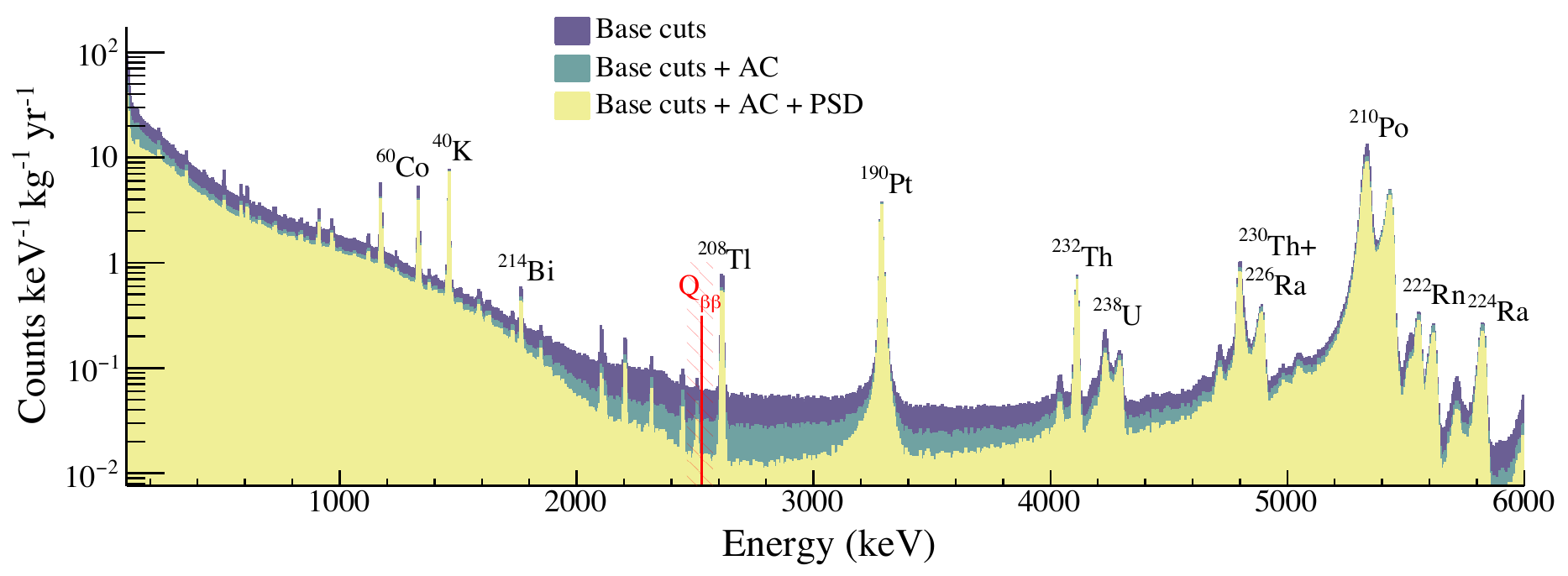}
\caption{\label{fig:spectrum}The CUORE energy spectrum for 2039.0~kg$\cdot$yr of TeO$_{2}$ exposure after each selection cut, the Q$_{\beta\beta}$ for $^{130}$Te is marked as a red line with shaded ROI. Figure reprinted from Ref.~\cite{cuoreScience2025}.}
\end{center}
\end{figure*}

The CUORE experiment has been taking data since 2017, more than 2.9~ton$\cdot$year raw TeO$_{2}$ exposure has been accumulated until now. After the offline data processing and several analysis cuts with a total efficiency of 93.4\%, the global energy spectrum of CUORE experiment with 2039.0~kg$\cdot$yr TeO$_{2}$ exposure is shown in Fig.~\ref{fig:spectrum}.
To perform the analysis in the ROI, the detector response to the monoenergetic energy deposition of $^{130}$Te 0$\nu\beta\beta$ decay has to be correctly evaluated. We firstly fit an empirical model (based on the calibration data) to the most prominent gamma peaks in physics data for each bolometer and dataset. Their energy resolutions and energy reconstruction bias, defined as the difference between the reconstructed and literature nominal peak energy, are hence determined. Then the scaling of both energy resolutions and bias are parameterized by quadratic polynomial functions of energy. The scaling functions are used to determine the detector response at Q$_{\beta\beta}$~\cite{cuore2022search, cuoreScience2025}. 

\begin{figure*}[!htb]
\begin{center}
\includegraphics[height=5.5cm]{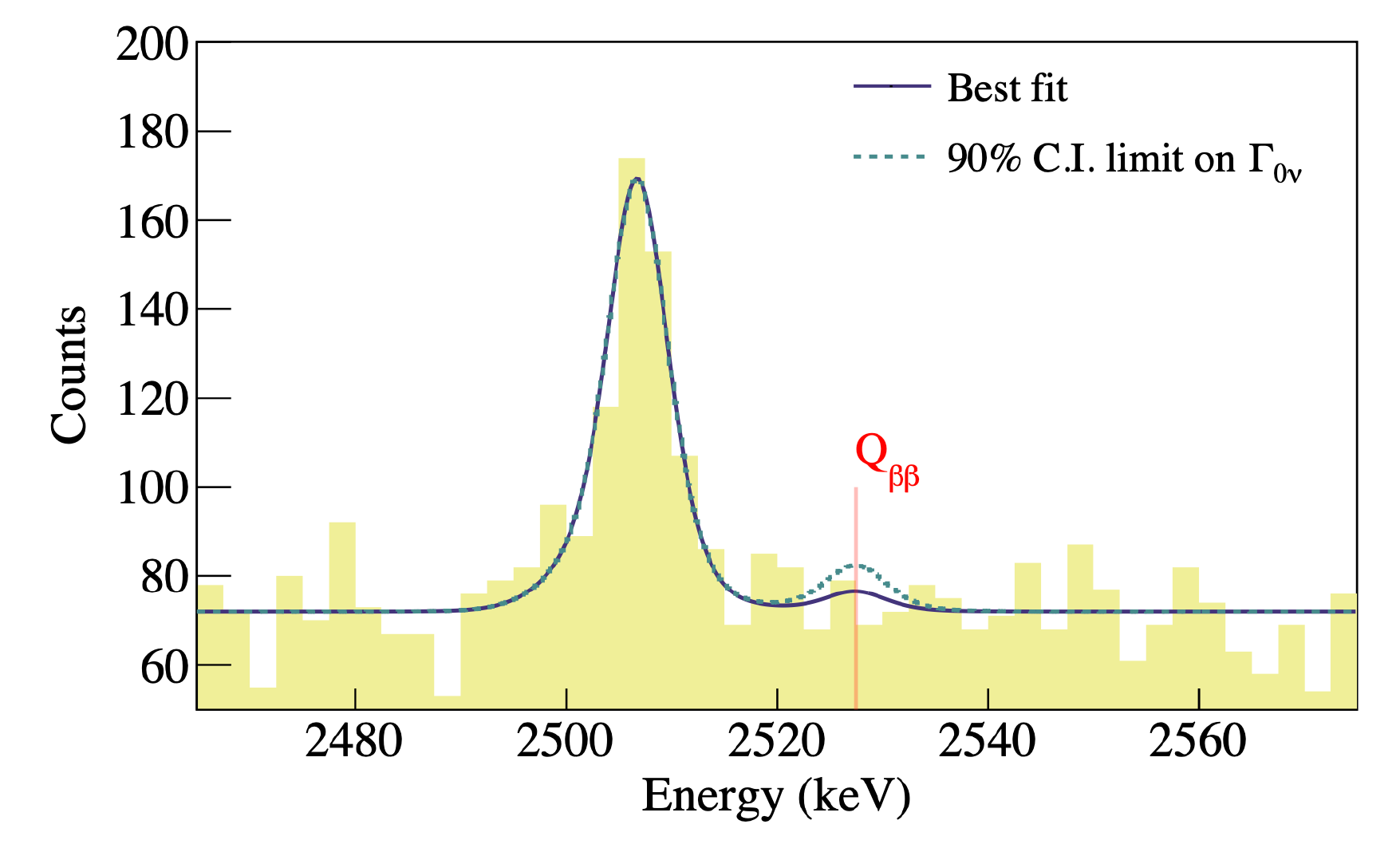}
\caption{\label{fig:roi_fit}\textbf{}Fit to the ROI spectrum. Figure reprinted from Ref.~\cite{cuoreScience2025}.}
\end{center}
\end{figure*}

Figure.~\ref{fig:roi_fit} shows the energy spectrum in the [2465, 2575]~keV fit region corresponding to 2039.0~kg$\cdot$yr of TeO$_{2}$ (567.0~kg$\cdot$yr of $^{130}$Te) exposure. The only one visible background peak at 2505.7~keV comes from the simultaneous absorption of two coincident $\gamma$-rays from $^{60}$Co in the same crystal. We perform an unbinned Bayesian fit to the spectrum using the Bayesian Analysis Toolkit (BAT)~\cite{CALDWELL20092197}, shown also in Fig~\ref{fig:roi_fit}. The spectrum is modeled as the sum of a Q$_{\beta\beta}$ peak, a dataset-dependent uniform background and a time-dependent $^{60}$Co sum peak. No statistically significant evidence of 0$\nu\beta\beta$ decay decay is observed with this analysis campaign. We come to a limit on the decay rate of $\Gamma^{0\nu}_{1/2} <2.0\times 10^{-26}$~yr$^{-1}$ at 90\% credibility interval (C.I.) corresponding to a decay half-life limit of $T^{0\nu}_{1/2}>3.5\times 10^{25}$ yr (90\% C.I.) for $^{130}$Te. We also extract an effective Majorana neutrino mass $m_{\beta\beta}$ based on the light-neutrino exchange model, giving $m_{\beta\beta}$<70–250~meV in accordance with different nuclear models~\cite{cuoreScience2025}. 
The CUORE experiment also obtains so far the most precise measurement of the $^{130}$Te 2$\nu\beta\beta$ half-life of $T^{2\nu}_{1/2}=(9.32^{+0.05}_{-0.04}\text{stat }^{+0.07}_{-0.07}\text{syst})\times10^{20}$~yr based on 1038~kg$\cdot$yr of collected exposure~\cite{jdhf-hn4l}, thanks to the robust background model built for COURE~\cite{PhysRevD.110.052003}.

\section{CUPID experiment}
The CUORE experiment has demonstrated the feasibility of applying large-scale cryogenic detectors for sensitive 0$\nu\beta\beta$ decay searching. Moreover, CUORE also provides critical prior knowledge for next-generation bolometric experiments. Its sensitivity is limited by $\alpha$ background originating from surface
contamination of the detector structure. 

The measurement based on phonon detection is equally sensitive to any kind of energy deposition. Exploiting another detection channel is essential to discriminate the background events from interested signals when using bolometers. CUPID, the CUORE Upgrade with Particle IDentification, will be housed in the CUORE cryostat, using scintillating bolometers for the search of 0$\nu\beta\beta$ decay of $^{100}$Mo. Two primary improvements are foreseen on the background suppression. One is the capability of simultaneously dual readout of heat and light allowing the discrimination of alpha particles from $\beta/\gamma$ particles, as their different light yield due to the quenching effect. The other is the higher Q-value 3034~keV of $^{100}$Mo, which lies in a ROI that free from natural radioactive background. Two pilot experiments CUPID-0 and CUPID-Mo have exhibited the effective background rejection capabilities and competitive sensitivities using cryogenic scintillating bolometers~\cite{PhysRevLett.129.111801, augier2022final}. 

The CUPID detector will consist of 1596 cubic scintillating Li$_{2}$MoO$_{4}$ crystals with side length of 45~mm, enriched to $\geq$95\% in $^{100}$Mo with a total $^{100}$Mo mass of 240~kg. 
As temperatures at the scale of 10~mK, the only possible light detectors (LDs) are also bolometers transferring the light collected into heat. The CUPID LDs consist of 300 $\mathrm{{\mu}}$m-thick high-purity germanium wafers coated with a 70 nm-thick silicon monoxide (SiO) anti-reflective layer to enhance the photon absorption. Other than that, a set of Al electrodes is instrumented on the LDs for high-voltage biasing to exploit the Neganov-Trofimov-Luke (NTL) effect~\cite{novati2019charge}. When photons interacts in the LD, phonons and electron-hole pairs are produced. The NTL effect works as drifting electron-hole pairs through an applied electric field between Al electrodes, generating additional phonons and amplify the thermal signal. Unlike the CUPID-0 and CUPID-Mo, the CUPID experiment will implement the light detection without reflective layers surrounding the crystals. Its absence will allow the identification of events induced by surface contaminants through the use of an anti-coincidence cut. The NTL effect plays a crucial role on the light signal amplification and pileup rejection, as it provides fast time response and a high signal-to-noise ratio. 

\begin{figure*}[!htb]
\centering
\includegraphics[height=6.5cm]{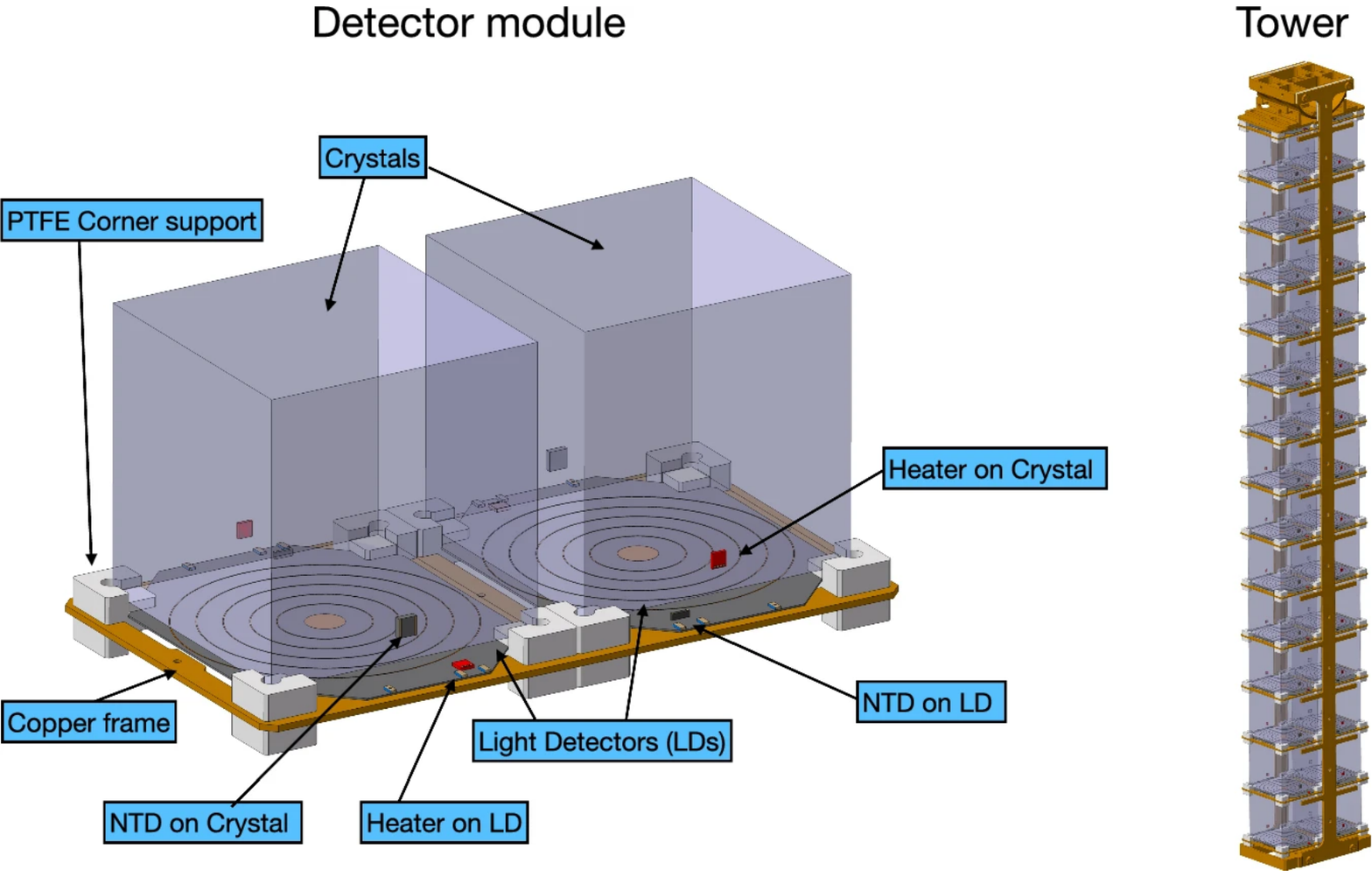}
\caption{Left: schematic view of a single CUPID floor with two side-by-side detector modules, each consisting of a Li$_{2}$MoO$_{4}$ absorber and a Ge light detector. Right: a single tower of 14 floors, or 28 detector modules. Figure reprinted from Ref.~\cite{cupid2025cupid}.\label{fig:cupidb}}
\end{figure*}
The CUPID baseline design foresees that the total 1596 crystals will be arranged in 57 towers of 14 floors, with two crystals per floor, as shown in Fig.~\ref{fig:cupidb}. The mechanical structure is modified with respect to that of CUORE, the structural parts are designed to allow them to be assembled without any screws, which is beneficial both for assembly and background control. The copper frames will be machining by laser cutting to simplify the subsequent cleaning process. A dedicated test campaign was carried out in 2021 at HallA of LNGS with a medium-scale prototype of 28 Li$_{2}$MoO$_{4}$ crystals and 30 Ge LDs. It has proved that a good energy resolution of $\sim$6.6~keV FWHM at 2615~keV and a sufficient light yield of 0.36~keV/MeV~\cite{cupid2025gravity}. In the present ongoing test campaign, all the light detectors are transformed into NTL light detectors, also a fiber is mounted along the tower for light detectors calibration. 

The CUPID experiment will undergo two stages: in the first stage, foreseen to start in 2030, 1/3 of the crystals will be deployed, while the second, full-scale stage is foreseen to commence in 2034. Based on the well-established background models of CUORE, CUPID-0 and CUPID-Mo experiments, the background goal of CUPID is set at 1.0$\times10^{-4}$~counts/(keV$\cdot$kg$\cdot$yr) in the ROI after a comprehensive evaluation~\cite{cupid2025cupid}. If the expected resolution of 5~keV FWHM at 3~MeV is achieved, with 10 years of livetime CUPID will reach a 90\% C.L. half-life exclusion sensitivity of 1.8$\times10^{27}$ yr, corresponding to a $m_{\beta\beta}$ sensitivity of 9$-$15~meV, and a 3$\sigma$ discovery sensitivity of 1$\times10^{27}$~yr, corresponding to a $m_{\beta\beta}$ range of 12$-$21~meV~\cite{cupid2025cupid}.
\begin{figure*}[!htb]
\centering
\includegraphics[height=6.5cm]{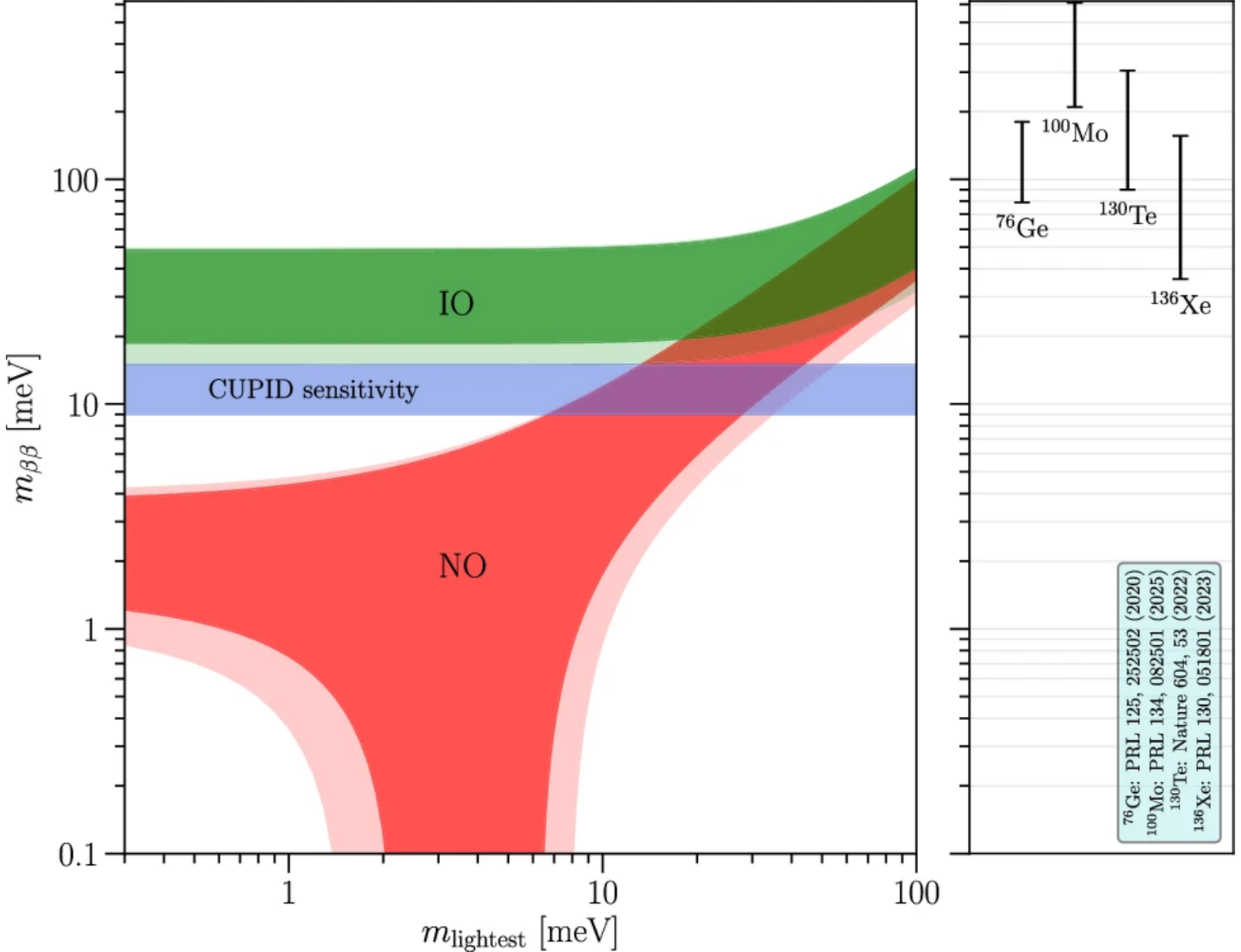}
\caption{Constraints on the effective Majorana neutrino mass ($m_{\beta\beta}$) as a function of the lightest neutrino mass ($m_{\rm{lightest}}$). Figure reprinted from Ref.~\cite{cupid2025cupid}.\label{fig:cupids}}
\end{figure*}

\section{Summary}
CUORE has demonstrated the scalability of the cryogenic calorimeter technique to tonne-scale experiments and its long-term operational stability, thereby paving the way for next-generation searches with cryogenic calorimeters. With 2039.0~kg$\cdot$yr TeO$_{2}$ exposure, we found no evidence of 0$\nu\beta\beta$ decay and set a new limit on the half life for such decay of $T^{0\nu}_{1/2}>3.5\times 10^{25}$ yr (90\%C.I.) for $^{130}$Te. CUORE will proceed taking data until reaching an analyzed exposure of 3~tonne$\cdot$yr TeO$_{2}$ ($\sim$1 tonne$\cdot$yr $^{130}$Te). The next-generation 0$\nu\beta\beta$ searching experiment CUPID will leverage the existing CUORE’s cryostat and infrastructure, performing the search for 0$\nu\beta\beta$ decay of $^{100}$Mo. With 10-years of data taking, CUPID will play a central role in the world-leading 0$\nu\beta\beta$ search and discovery program.

\acknowledgments
The CUORE Collaboration thanks the directors and staff of the Laboratori Nazionali
del Gran Sasso and the technical staff of our laboratories. This work was supported
by the Istituto Nazionale di Fisica Nucleare (INFN); the National Science Foundation
under Grant Nos. NSF-PHY-0605119, NSF-PHY-0500337, NSF-PHY-0855314, NSF-PHY-0902171, NSF-PHY-0969852, NSF-PHY-1307204, NSF-PHY-1314881, NSF-PHY-1401832, NSF-PHY-1913374, and NSF-PHY-2412377; Yale University, Johns Hopkins University, and University of Pittsburgh. This material is also based upon work supported by the US Department of Energy (DOE) Office of Science under Contract Nos. DE-AC02-05CH11231, and DE-AC52-07NA27344; by the DOE Office of Science, Office of Nuclear Physics under Contract Nos. DE-FG02-08ER41551, DE-FG03-00ER41138, DE-SC0012654, DE-SC0020423, DE-SC0019316, and DE-SC0011091. This research used resources of the National Energy Research Scientific Computing Center (NERSC). This work makes use of both the DIANA data analysis and APOLLO data acquisition software packages, which were developed by the CUORICINO, CUORE, LUCIFER, and CUPID-0 Collaborations. The author acknowledges the Advanced Research Computing at Virginia Tech and the Yale Center for Research Computing for providing computational resources and technical support that have contributed to the results reported within this work.

The CUPID Collaboration thanks the directors and staff of the Laboratori Nazionali del Gran Sasso and the technical staff of our laboratories. This work was supported by the Istituto Nazionale di Fisica Nucleare (INFN); by the European Research
Council (ERC) under the European Union Horizon 2020 program (H2020/2014-2020) with the ERC Advanced Grant no. 742345 (ERC-2016-ADG, project CROSS) and the Marie Sklodowska-Curie Grant Agreement No. 754496; by the Italian Ministry of University and Research (MIUR) through the grant Progetti di ricerca di Rilevante Interesse Nazionale (PRIN) grant no. 2017FJZMCJ and grant no. 2020H5L338; by the US National Science Foundation under Grant Nos. NSF-PHY-1401832, NSF-PHY-1614611, NSF-PHY-2412377 and NSF-PHY-1913374; by the French Agence Nationale de la Recherche
(ANR) through the ANR-21-CE31-0014- CUPID-1; by the National Research Foundation of Ukraine (Grant No. 2023.03/0213). This material is also based upon work supported by the US Department of Energy (DOE) Office of Science under Contract Nos. DE-AC02-05CH11231 and DE-AC02-06CH11357; and by the DOE Office of Science, Office
of Nuclear Physics under Contract Nos. DE-FG02-08ER41551, DE-SC0011091, DE-SC0012654, DE-SC0019316, DE-SC0019368, and DE-SC0020423. This work was also supported by the Russian Science Foundation under grant No. 18-12-00003. This research used resources of the National Energy Research Scientific Computing Center
(NERSC). This work makes use of both the DIANA data analysis and APOLLO data acquisition software packages, which were developed by the CUORICINO, CUORE, LUCIFER and CUPID-0 Collaborations.




\bibliographystyle{JHEP}
\bibliography{biblio.bib}

\end{document}